\documentclass[aps,pra,showpacs,floatfix,superscriptaddress,twocolumn,bibnotes]{revtex4-1}
\usepackage{graphicx}
\usepackage{amsmath}
\usepackage{bm}
\usepackage{color}
\usepackage{dcolumn}
\usepackage{hyperref}

\makeatletter
\newcommand*{\balancecolsandclearpage}{%
  \close@column@grid
  \clearpage
  \twocolumngrid
}
\makeatother

\begin{document}

\newcommand{\etal}{{\it et al.}}
\newcommand{\pprime}{{\prime\prime}}


\newcommand{\NIST}{
National Institute of Standards and Technology, 325 Broadway, Boulder, Colorado 80305, USA}

\newcommand{\CU}{
Department of Physics, University of Colorado,
Boulder, Colorado 80309, USA}

\newcommand{\Korea}{
Department of Physics, Korea University, Seoul 02841, Republic of Korea
}

\newcommand{\GTRI}{
Georgia Tech Research Institute, Atlanta, GA 30332, USA
}

\newcommand{\StableLasers}{
Stable Laser Systems, Boulder, Colorado 80301, USA
}

\newcommand{\China}{
State Key Laboratory of Advanced
Optical Communication Systems and Networks, Institute
of Quantum Electronics, School of Electronics Engineering
and Computer Science, Peking University, Beijing
100871, China
}

\newcommand{\NBI}{
Niels Bohr Institute, University of Copenhagen, Blegdamsvej 17, 2100 Copenhagen, Denmark
}

\title{Faraday-shielded, DC Stark-free optical lattice clock}


\author{K. Beloy}
\altaffiliation{kyle.beloy@nist.gov}
\affiliation{\NIST}
\author{X. Zhang}
\altaffiliation[permanent address: ]{\China}
\affiliation{\NIST}
\author{W. F. McGrew}
\affiliation{\NIST}
\affiliation{\CU}
\author{N. Hinkley}
\altaffiliation[present address: ]{\StableLasers}
\affiliation{\NIST}
\affiliation{\CU}
\author{T. H. Yoon}
\altaffiliation[permanent address: ]{\Korea}
\affiliation{\NIST}
\author{D. Nicolodi}
\affiliation{\NIST}
\author{R. J. Fasano}
\affiliation{\NIST}
\affiliation{\CU}
\author{S. A. Sch\"{a}ffer}
\altaffiliation[permanent address: ]{\NBI}
\affiliation{\NIST}
\author{R. C. Brown}
\altaffiliation[present address: ]{\GTRI}
\affiliation{\NIST}
\author{A. D. Ludlow}
\affiliation{\NIST}

\date{\today}

\begin{abstract}
We demonstrate the absence of a DC Stark shift in an ytterbium optical lattice clock. Stray electric fields are suppressed through the introduction of an in-vacuum Faraday shield. Still, the effectiveness of the shielding must be experimentally assessed. Such diagnostics are accomplished by applying high voltage to six electrodes, which are grounded in normal operation to form part of the Faraday shield. Our measurements place a constraint on the DC Stark shift at the $10^{-20}$ level, in units of the clock frequency. Moreover, we discuss a potential source of error in strategies to precisely measure or cancel non-zero DC Stark shifts, attributed to field gradients coupled with the finite spatial extent of the lattice-trapped atoms. With this consideration, we find that Faraday shielding, complemented with experimental validation, provides both a practically appealing and effective solution to the problem of DC Stark shifts in optical lattice clocks.
\end{abstract}

\maketitle

In the nearly seven decade-old quest to push the boundaries of atomic clock performance, and thus metrological capabilities in general, the elimination or precise evaluation of frequency shifts caused by external electromagnetic fields has been a persistent challenge \cite{Har52,EssPar55}. In modern-day optical lattice clocks, AC Stark shifts due to lattice light and blackbody radiation are two prominent examples \cite{KatTakPal03,LudBoyYe15}. DC Stark shifts, attributed to nearby electronics or patch charges on the clock apparatus, have been observed as large as $10^{-13}$ \cite{LodZawLor12} and pose a legitimate threat to state-of-the-art $10^{-18}$ clock performance (throughout, quoted shifts are understood to be in units of the clock frequency). Strategies to mitigate this threat include I) applying electric fields to measure and, if desired, cancel the stray-field shift~\cite{LodZawLor12,FalLemGre14,BloNicWil14,NicCamHut15,NorCliMun17arXiv}, or II) enclosing the atoms by equipotential conductive surfaces, furnishing them with a field-free environment \cite{BelHinPhi14,UshTakDas14,NemOhkTak16,KolGroVog17}. DC Stark shifts have also been estimated from apparatus geometry and material properties~\cite{PizThoRau17,KimHeoLee17}. Recently, Rydberg atoms were demonstrated as an {\it in situ} probe of the stray field in an optical lattice clock~\cite{BowHobHui17}.

Here we identify a mechanism capable of compromising a Method I analysis, for which uncertainties at the $10^{-19}$ level have been reported. Using a simple model, we demonstrate how field gradients coupled with finite spatial extent of the lattice-trapped atoms can lead to appreciable clock error. Generally, the error scales with the measured stray-field shift. In principle, such error can be reduced by minimizing the stray field itself, which is precisely the objective of Method II. Unfortunately, practical constraints preclude surrounding the atoms with an ideal, continuous Faraday cage. Moreover, even conductive surfaces can acquire patch charges, a known concern for electrodes in ion clocks~\cite{BerMilBer98}. Consequently, a residual shift may remain, and quantifying an upper bound may be challenging. Seemingly, an optimal solution combines the attributes of Methods I and II. We demonstrate this combined approach in an ytterbium optical lattice clock, with measurements confirming the absence of a stray-field shift at the $10^{-20}$ level.

Given a uniform static electric field $\mathbf{E}$, the clock acquires a frequency shift $\delta\nu=kE^2$,
where $E=\left|\mathbf{E}\right|$ and $k$ is specific to the clock transition. Namely, $k\equiv-(\alpha_e-\alpha_g)/2h$, where $h$ is Planck's constant and $\alpha_{g,e}$ are the static polarizabilities of the ground and excited clock states. To characterize blackbody radiation shifts, the coefficient $k$ has been accurately measured for both Yb and Sr clock transitions \cite{SheLemHin12,MidFalLis12}.

In practice, the lattice-trapped atoms have finite spatial extent, and the electric field may be nonuniform over this extent. Thus, a more complete representation of the clock shift is $\delta\nu=k\left\langle E^2\right\rangle$,
where $\langle\cdots\rangle$ denotes an average over the atoms. Generally, $\mathbf{E}$ is composed of both stray and applied fields. Given some nonzero stray field, it is evident that a true null shift can only be achieved if the applied field identically cancels the stray field across the entire atomic extent.

\begin{figure}[t]
\includegraphics[width=\linewidth]{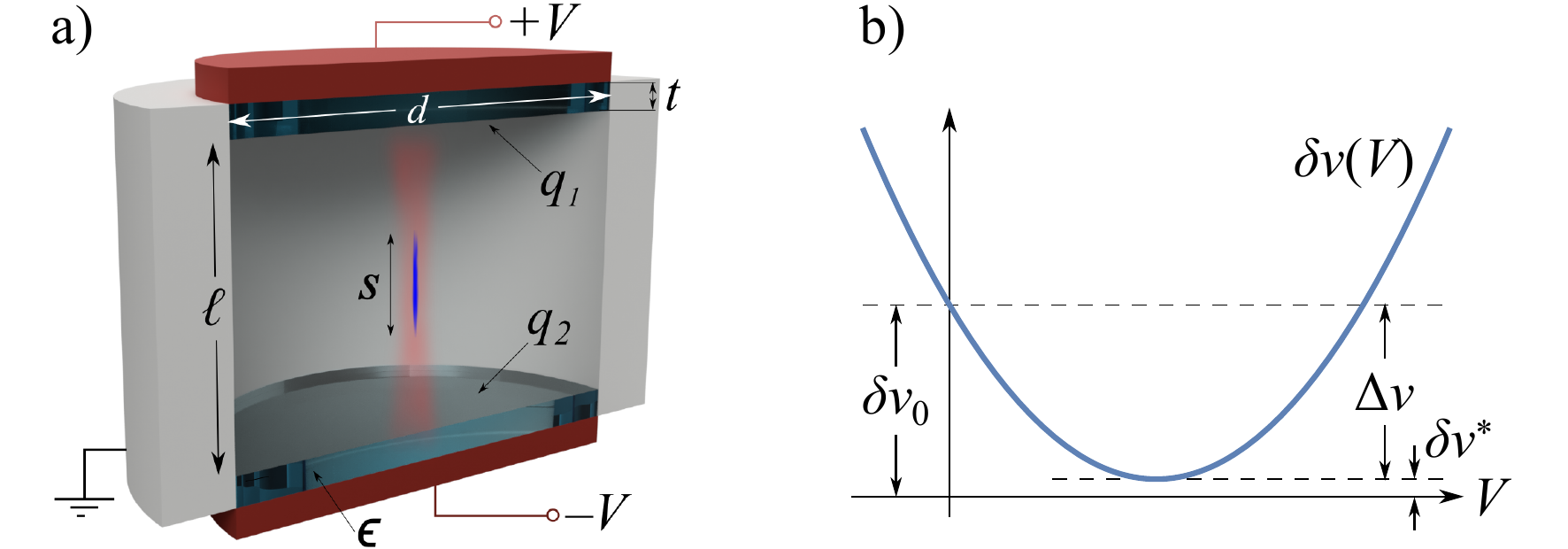}
\caption{a) Section view of the clock model described in the text. b) Corresponding clock shift $\delta\nu(V)$, with the quantities $\delta\nu_0$, $\delta\nu^*$, and $\Delta\nu$ introduced in the text. For $k>0$ ($k<0$), the extremum is a minimum (maximum) and all three quantities are positive (negative). 
}
\label{Fig:modelquadcurve}
\end{figure}

To illustrate the role field gradients can play in Method I, we introduce a simple model that affords an analytical solution. The model is illustrated in Fig.~\ref{Fig:modelquadcurve}(a) and amounts to a cylindrically symmetric boundary value problem for the fields. The vacuum apparatus is taken to be a hollow metallic cylinder sealed with glass windows. The cylinder is electrically grounded, while the windows carry uniformly distributed static charges $q_1$ and $q_2$ across their respective internal surfaces. The external surfaces are spanned by electrodes, to which opposite voltages $+V$ and $-V$ are applied. With the electrodes grounded ($V=0$), a stray field exists due to the charges. For $V\neq0$, the electrodes further introduce an applied field. A one-dimensional optical lattice aligned with the symmetry axis confines the atoms with negligible radial extent and Gaussian axial distribution $(2\pi s^2)^{-1/2}\exp\left(-z^2/2s^2\right)$, with $z$ being the distance from the center of the vacuum apparatus. The windows are separated by a distance $\ell$ and have diameter $d$, thickness $t$, and dielectric constant $\epsilon$. Expressions for the electric potential within the vacuum region can be found in the Supplemental Material (SM) \cite{SM}.


As demonstrated on a more general basis below, the clock shift has the functional form
\begin{equation}
\delta\nu\left(V\right)=\delta\nu_0+aV+bV^2.
\label{Eq:dnuoneV}
\end{equation}
The coefficients $a$ and $b$ are experimentally accessible parameters whose values may be determined by modulating $V$ and observing the clock response. Specifying the clock shift for any $V$ requires further knowledge of the stray-field shift $\delta\nu_0$. Towards this goal, we consider the extremum value of $\delta\nu\left(V\right)$, denoted $\delta\nu^*$. The stray-field shift $\delta\nu_0$, the extremum shift $\delta\nu^*$, and the difference between them $\Delta\nu\equiv \delta\nu_0-\delta\nu^*$ are depicted in Fig.~\ref{Fig:modelquadcurve}(b). In contrast to $\delta\nu_0$ and $\delta\nu^*$, $\Delta\nu$ is accessible through modulation of $V$. Invoking elementary calculus with Eq.~(\ref{Eq:dnuoneV}), we find $\Delta\nu=a^2/4b$. 

Let us initially neglect the atomic extent, taking the limit $s\rightarrow0$. In this case, there exists a $V$ for which the applied field identically cancels the stray field at the atoms, resulting in a null clock shift. This necessarily coincides with the extremum of $\delta\nu\left(V\right)$, as any other $V$ yields a nonzero clock shift of definite sign (determined by $k$). This implies $\delta\nu^*=0$, and it follows that $\delta\nu_0$ may be inferred from $\Delta\nu$ according to $\delta\nu_0=\Delta\nu$.

The above reasoning breaks down for nonzero $s$, as we can no longer expect there to be a $V$ such that the applied field identically cancels the stray field over the entire atomic extent. Consequently, $\delta\nu^*$ plays the role of a frequency correction for the field gradients. We write $\delta\nu^*=\eta\Delta\nu$, motivated by the fact that $\delta\nu^*$ and $\Delta\nu$ scale similarly with the stray field. Namely, a uniform scaling of the stray charge leaves $\eta$ unchanged. The stray-field shift subsequently reads $\delta\nu_0=\left(1+\eta\right)\Delta\nu$. To leading order in $s$, we find $\eta=\zeta^2s^2/\mathcal{R}^2$,
where $\mathcal{R}$ is an effective length whose expression is given in the SM \cite{SM} and $\zeta\equiv(q_1+q_2)/(q_1-q_2)$ quantifies the charge-symmetry between the windows. Choosing $d=150$~mm, $\ell=100$~mm, $t=10$~mm, and $\epsilon=3.8$, $\mathcal{R}$ evaluates to $\mathcal{R}=42$~mm. Further assuming $s=1$~mm and 25\% more charge on one window than the other, we obtain $\eta\approx0.05$.

The example above suggests that the frequency correction $\delta\nu^*$ may be non-negligible if $\delta\nu_0$ itself is appreciable. More specifically, for an optical lattice clock exhibiting a stray-field shift above $10^{-18}$, specification or cancellation of the shift at or below $10^{-18}$ may not be straightforward using Method I. There exists an additional burden in quantifying this correction or validating its neglect. Moreover, this effect could influence evaluations of other systematic effects. For example, lattice light shifts are typically characterized by varying lattice intensity, which may vary the atomic extent. Lastly, while our model suffices to demonstrate the potential importance of this effect, an actual system will inevitably be more complicated (lack symmetry in the apparatus, charge distribution, and atomic distribution; be an open-boundary system for the fields; contain dielectric surfaces in close proximity to the atoms; etc.). These complexities will presumably add to the difficulty of quantifying the correction due to field gradients.

From the preceding discussion, there is clear motivation for minimizing the stray-field shift. Recently, our group demonstrated an in-vacuum ``shield'' surrounding the lattice-trapped atoms in a Yb optical lattice clock~\cite{BelHinPhi14}. An updated version is pictured in Fig.~\ref{Fig:shield}. The shield's objective is twofold: provide a well-defined, near-ideal room-temperature blackbody radiation environment and suppress stray electric fields. The shield body is a single copper structure, internally coated with electrically-conductive carbon nanotubes. BK7 windows for optical access have an electrically-conductive, $\sim\!5$ nm-thick indium tin oxide (ITO) based coating. With the exception of two small apertures for atomic access, the copper body and ITO-coated windows collectively enclose the atoms. 
The design facilitates a Method I analysis. The windows are electrically isolated from the copper body using thin silicone spacers while being electrically connected to an external high-voltage source. We have independent control of the voltage on six windows, constituting three opposing pairs nominally aligned along mutually orthogonal axes. 
A seventh window is electrically connected to the copper body, which is permanently grounded. 

\begin{figure}[t]
\includegraphics[width=\linewidth]{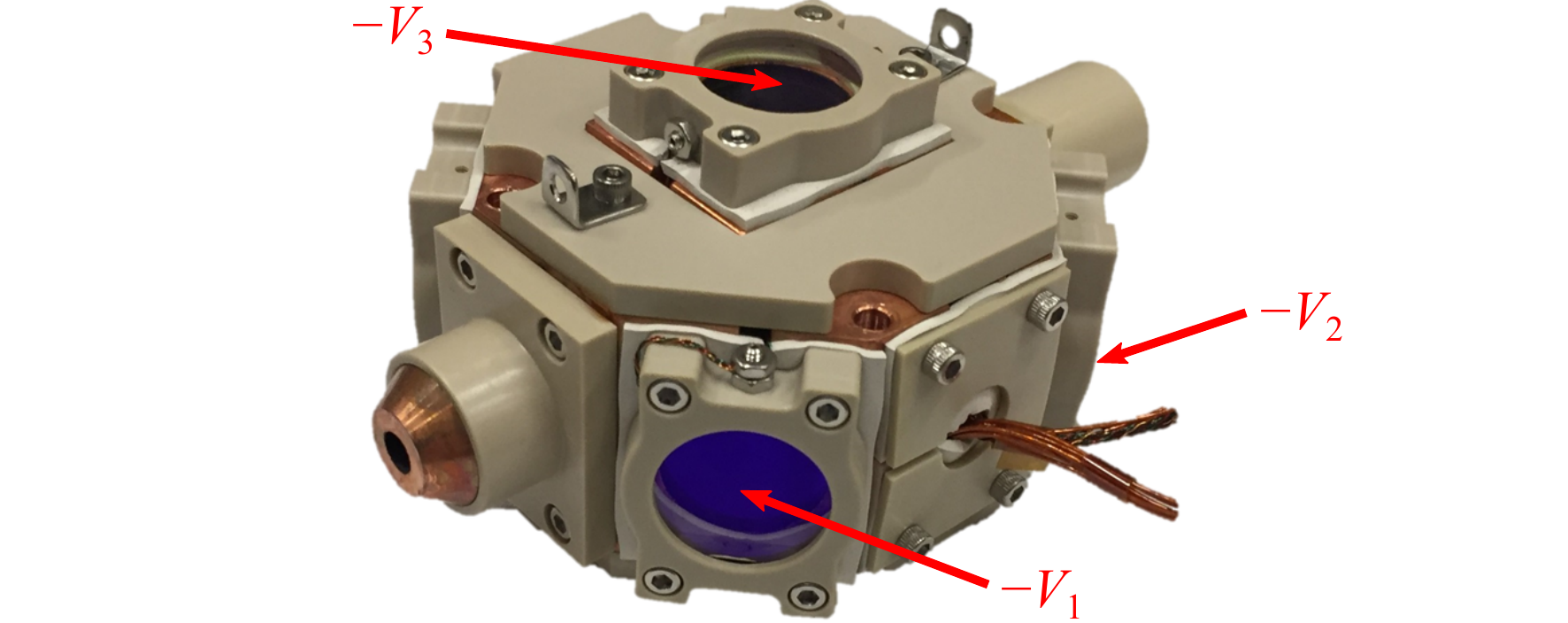}
\caption{Faraday shield described in the text, with ITO coated windows. Voltages $-V_1$, $-V_2$, and $-V_3$ are assigned to the indicated windows, with voltages $+V_1$, $+V_2$, and $+V_3$ assigned to the respective opposing windows (unlabeled). The shield body is a single copper structure, internally coated with carbon nanotubes. PEEK plastic secures the windows to the body, suppresses radiative heat exchange with the environment, and hides functional components including electrical wires, resistance temperature detectors, and film heaters. For loading the optical lattice, a thermal beam of atoms enters the shield through an aperture (pictured bottom left), while a counterpropagating beam of slowing light enters through an opposing aperture. Bundled electrical wires (pictured bottom right) proceed to vacuum feedthroughs. For scale, the windows are 1 inch in diameter.}
\label{Fig:shield}
\end{figure}

The model introduced above, Fig.~\ref{Fig:modelquadcurve}(a), is quasi-one-dimensional (we refer to it as the 1D model below). That is, although the stray and applied fields are derived from a three-dimensional boundary value problem, only the symmetry axis is sampled by the atoms, with the symmetry ensuring that the fields align (or anti-align) along this axis. Practical clocks, such as ours, demand a more general theory. To this end, we allow multiple voltage variables $V_i$. We assign to each electrode in the system a linear combination of the $V_i$, defining the voltage applied to that electrode. 
The total electric field is $\mathbf{E}_0+\sum_i\mathbf{E}_i$, where $\mathbf{E}_0$ is the stray field (assumed independent of the $V_i$) and $\sum_i\mathbf{E}_i$ is the applied field with $\mathbf{E}_i\propto V_i$. The clock shift subsequently reads
\begin{equation}
\delta\nu\left(V_1,V_2,\cdots\right)=\delta\nu_0+\sum_ia_iV_i+\sum_{ij}b_{ij}V_iV_j,
\label{Eq:dnumultiV}
\end{equation}
where $\delta\nu_0=k\left\langle E_0^2\right\rangle$, $a_iV_i=2k\left\langle \mathbf{E}_0\cdot\mathbf{E}_i\right\rangle$, and $b_{ij}V_iV_j=k\left\langle \mathbf{E}_i\cdot\mathbf{E}_j\right\rangle$. 
Equation (\ref{Eq:dnumultiV}) is a generalization of equation~(\ref{Eq:dnuoneV}) above. 
As before, we introduce the difference $\Delta\nu\equiv\delta\nu_0-\delta\nu^*$, where $\delta\nu^*$ is the extremum with respect to all $V_i$. In terms of the coefficients in Eq.~(\ref{Eq:dnumultiV}), $\Delta\nu$ reads~\cite{myfootnote}
\begin{equation}
\Delta\nu=\frac{1}{4}\sum_{ij}a_ia_j\left(b^{-1}\right)_{ij},
\label{Eq:DeltanumultiV}
\end{equation}
where the $\left(b^{-1}\right)_{ij}$ are related to the $b_{ij}$ through matrix inversion (regarding the latter as elements of a matrix $b$ and the former as elements of its inverse $b^{-1}$). We partition the problem of specifying $\delta\nu_0$ into two parts according to the sum $\delta\nu_0=\Delta\nu+\delta\nu^*$. We initially focus on $\Delta\nu$, which represents the portion accessible through modulation of the $V_i$.



For our shield, we introduce a voltage variable for each opposing window pair, with voltages $-V_i$ and $+V_i$ assigned to the windows of each pair $i=1,2,3$ (refer to Fig.~\ref{Fig:shield}). Given these assignments, the corresponding $\mathbf{E}_i$ are nominally uniform and mutually orthogonal at the atoms. In the limit this is strictly true, the $b_{ij}$ with $i\neq j$ vanish and Eq.~(\ref{Eq:DeltanumultiV}) reduces to
\begin{equation}
\Delta\nu=\sum_i\frac{a_i^2}{4b_{ii}},
\label{Eq:Deltanusimple}
\end{equation}
indicating that contributions to $\Delta\nu$ can be evaluated independently for each ``direction'' and summed up. However, nonorthogonality or nonuniformity of the $\mathbf{E}_i$ must be considered. This could be assessed through independent means, such as geometrical considerations and field modeling. A more reliable estimate exploits the atoms themselves. Either way, once assessed, these effects can be treated perturbatively, as highlighted in the SM~\cite{SM}.


Alternatively, without claiming {\it a priori} knowledge about the $\mathbf{E}_i$, here we use the general Eq.~(\ref{Eq:DeltanumultiV}). We reserve Eq.~(\ref{Eq:Deltanusimple}) for future DC Stark shift assessments, where data from the present work can be leveraged to improve measurement efficiency and constrain deviations to this simple expression. We measure the induced shift $\delta\nu\left(V_1,V_2,V_3\right)-\delta\nu_0$ for various combinations of the arguments $V_i$. The $V_i$ define a ``test'' configuration, with the fully-grounded configuration serving as a common reference. Each measurement run involves interleaving interrogations for the two configurations (test and reference) and recording the frequency difference. Table~\ref{Tab:results} presents our data. We ascribe to each measurement a statistical uncertainty commensurate with the Allan deviation at the end of the run, $\sim\!1\times10^{-17}$. Voltage switching is enacted on the millisecond timescale, with spectroscopy initiated a few hundred milliseconds afterwards. Applied voltages are assessed with a voltage divider and found to be well-defined at the $2\times10^{-4}$ fractional level. This introduces negligible uncertainty, with the exception of one line in Table~\ref{Tab:results}. For this data, large shifts were induced, $-9\times10^{-14}$, with an uncertainty principally due to the applied voltage. All other induced shifts are at the $10^{-16}$ level. For each combination of $V_i$, measurements were performed under opposite polarity conditions. 


\newcommand{\e}[1]{\times10^{#1}}
\newcommand{\tbl}{5mm}
\newcommand{\tbox}[3]{(\makebox[\tbl][c]{#1},\makebox[\tbl][c]{#2},\makebox[\tbl][c]{#3})}
\newcommand{\fnm}[1]{\footnotemark[#1]}
\newcommand{\fup}{f_\uparrow~\left(\e{-16}\right)}
\newcommand{\fdn}{f_\downarrow~\left(\e{-16}\right)}
\begin{table}[t]
\caption{Induced frequency shifts relative to the fully-grounded arrangement,~$\delta\nu\left(V_1,V_2,V_3\right)-\delta\nu_0$. Nonzero $V_i$ are specified by sign only, with $\left|V_1\right|=\left|V_2\right|=2$~kV and $\left|V_3\right|=110$~V except where noted. Left and right data columns correspond to upper and lower signs in the voltage specifications and represent opposite polarity conditions.}
\label{Tab:results}
\begin{ruledtabular}
\begin{tabular}{lcc}
\tbox{$V_1$}{$V_2$}{$V_3$}		& \multicolumn{2}{c}{induced shift $\left(\e{-16}\right)$}		\\
\hline\vspace{-2.5mm}\\
\tbox{$\pm$}{0}{0}				& $-2.81(10)$		& $-2.78(10)$	\\
\tbox{0}{$\pm$}{0}				& $-2.76(10)$		& $-2.76(9)$	\\
\tbox{0}{0}{$\pm$}				& $-2.77(7)$\fnm{1}	& $-2.67(7)$\fnm{1}	\\
\tbox{0}{0}{$\pm$}\fnm{2}		& $-915.2(4)$		& $-915.6(4)$	\\
\tbox{$\pm$}{$\pm$}{0}			& $-6.21(9)$		& $-6.09(9)$	\\
\tbox{$\pm$}{$\mp$}{0}			& $-5.15(9)$		& $-5.09(10)$	\\
\tbox{$\pm$}{0}{$\pm$}			& $-5.82(10)$		& $-5.92(11)$	\\
\tbox{$\pm$}{0}{$\mp$}			& $-5.60(10)$		& $-5.55(11)$	\\
\tbox{0}{$\pm$}{$\pm$}			& $-5.83(10)$		& $-5.94(10)$	\\
\tbox{0}{$\pm$}{$\mp$}			& $-5.49(10)$		& $-5.41(10)$	
\end{tabular}
\footnotetext[1]{weighted mean of two measurement runs}
\footnotetext[2]{$\left|V_3\right|=2$~kV}
\end{ruledtabular}
\end{table}

\begin{figure}[tb]
\includegraphics[width=\linewidth]{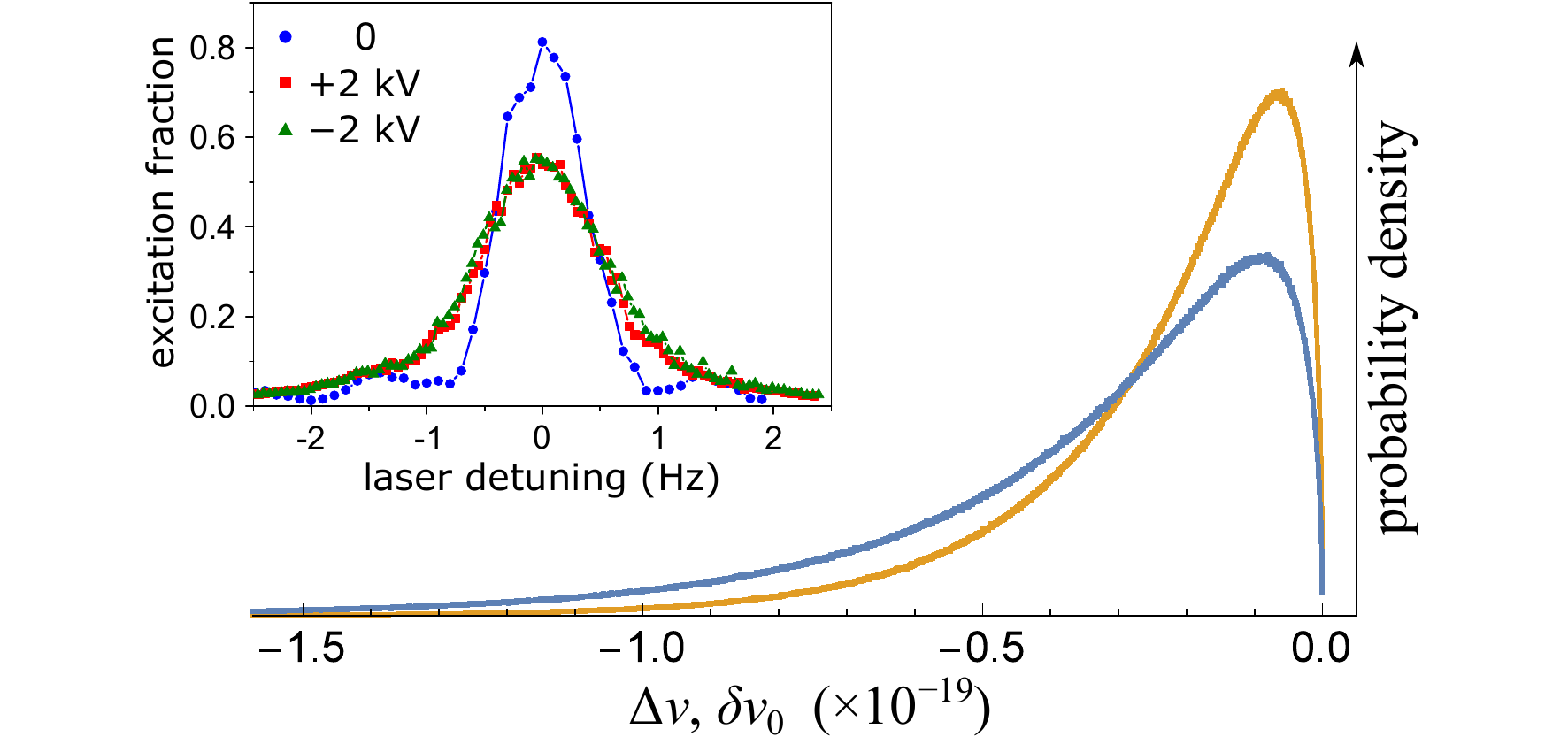}
\caption{Main: Probability distributions for $\Delta\nu$ (yellow curve) and $\delta\nu_0$ (blue curve). The quantities are related by $\delta\nu_0=\left(1+\eta\right)\Delta\nu$; the $\delta\nu_0$ distribution is broadened relative to the $\Delta\nu$ distribution due to uncertainty from $\eta$ (see text). Visible ``noise'' stems from the Monte Carlo evaluation. Inset: Excitation fraction versus laser detuning for the clock transition, without ($V_3=0$) and with ($V_3=\pm2$~kV) an applied field. $V_1=V_2=0$ in each case. Detuning is referenced from the respective linecenter. A polarity-independent broadening accompanies the applied field.}
\label{Fig:distrospectra}
\end{figure}

A cursory examination of Table~\ref{Tab:results} reveals no statistically-significant difference under any polarity reversal. This invariance immediately suggests $\Delta\nu\approx0$, though a more definitive analysis is clearly desired. 
Fitting Eq.~(\ref{Eq:dnumultiV}) to the data in Table~\ref{Tab:results} allows determination of the coefficients $a_i$ and $b_{ij}$, which can then be used to find $\Delta\nu$ via Eq.~(\ref{Eq:DeltanumultiV}).
We implement a Monte Carlo protocol~\cite{SM} to map probability distributions for the data (interpreted as uncorrelated Gaussian distributions) into a probability distribution for $\Delta\nu$, the result of which is presented in Fig.~\ref{Fig:distrospectra}. 
The distribution is clearly non-Gaussian and effectively constrains $\Delta\nu$ to negative values. The sign constraint is not surprising, considering all induced shifts are well-resolved negative (indicating $k<0$ with near certainty, in agreement with the known value~\cite{SheLemHin12}). Based on this distribution, we assert a 68.3\% confidence interval $-2.8\times10^{-20}<\Delta\nu<0$ and a 95.5\% confidence interval $-6.7\times10^{-20}<\Delta\nu<0$.

In order to specify $\delta\nu_0$, we must further address $\delta\nu^*$. For any point near the shield's center, 
an arbitrary applied field can, in principle, be constructed with an appropriate choice of the $V_i$. It is therefore possible to cancel an arbitrary stray field at that point. However, it is generally not possible to cancel the stray field over some extended volume. In complete analogy to the 1D model, $\delta\nu^*$ plays the role of a correction for field gradients.

To explore the role of gradients in our clock, we consider the atomic spectra in the Fig.~\ref{Fig:distrospectra} inset, obtained with one-second-long Rabi excitation.  The blue trace shows the spectrum when all electrodes are grounded (i.e., normal operation), yielding a $\sim\!1$~Hz Fourier-limited linewidth. The red trace shows the spectrum with a large applied field ($V_1=V_2=0$, $V_3=+2$~kV). Because the lattice-trapped atoms are 1--2 mm from the symmetry plane between the charged windows, they experience a linear gradient from the applied field, resulting in the observed inhomogeneous broadening (with broadening attributed to noise in the applied voltage smaller by an order of magnitude). The magnitude of this gradient is corroborated by finite element analysis~\cite{comsol}.  Were a stray-field gradient also present, it could add to the applied-field gradient and the observed line broadening.  Under opposite polarity conditions ($V_3=-2$~kV), the applied field and its gradient reverse.  In this case, the stray-field gradient would subtract from the applied-field gradient, reducing the observed line broadening.  The green trace shows the observed spectrum upon polarity reversal.  Since this reversal does not change the observed spectral linewidth and amplitude, a bound can be placed on the stray-field gradient.  This technique benefits from the large applied field at the atoms, which amplifies broadening from a stray-field gradient. For this axis of measurement, the constraint is $\left|\delta\nu^*\right|<2\times10^{-20}$.

While this technique could be repeated along the transverse lattice axes (where atomic extent is smaller and thus less sensitive to gradients) to constrain $\delta\nu^*$, here we exploit the fact that $\Delta\nu$ is essentially zero.  As done previously for the 1D model, we write $\delta\nu^*=\eta\Delta\nu$, motivated by the fact that $\delta\nu^*$ and $\Delta\nu$ scale similarly with the stray field. 
While we lack a means to precisely evaluate $\eta$, the need is alleviated by our tight constraint on $\Delta\nu$. To investigate plausible values of $\eta$, we perform a finite element analysis of our shield plus atoms. Stray fields are introduced by applying patch voltages on internal shield surfaces (here $\eta$ is unaffected by a uniform scaling of the patch voltages). Within the physical constraint $\eta\geq0$, arbitrary values of $\eta$ can be manufactured. Larger values require increasingly fine-tuned conditions. To realize $\eta>1$, for instance, a high degree of symmetry is required between patch voltages on opposing sides of the shield; given a sufficiently symmetric arrangement, the atoms must then reside at a precise location. By examining various conditions, we take $\eta=1$ as a conservative upper limit for our clock. A smaller value could be argued, but there is little incentive to be more aggressive or meticulous in light of our tight constraint on $\Delta\nu$.

Finally, we assume complete ignorance of $\eta$ between zero and unity, assigning it a uniform probability distribution over this range. Combined with our results for $\Delta\nu$, we derive a probability distribution for $\delta\nu_0=\left(1+\eta\right)\Delta\nu$. Figure~\ref{Fig:distrospectra} presents the distribution, from which we assert a 68.3\% confidence interval $-4.1\times10^{-20}<\delta\nu_0<0$ and a 95.5\% confidence interval $-1.0\times10^{-19}<\delta\nu_0<0$. For comparison, the largest systematic uncertainties in our clock are presently at the $1\times10^{-18}$ level~\cite{BelHinPhi14,BroPhiBel17}.

In conclusion, we have implemented Faraday shielding in an optical lattice clock and have constrained the stray-field DC Stark shift to below $10^{-19}$. In contrast to optical lattice clocks that lack Faraday shielding and exhibit nonzero stray-field shifts, our normal operation does not require regular spectroscopic monitoring of the shift, and there is no compromise to clock stability. Further measurements not part of this analysis, dispersed over multiple months and performed on independent Faraday-shielded clocks, have always yielded results consistent with zero stray-field shift. Here we have also identified a potential source of error in the measurement or cancellation of nonzero stray-field shifts attributed to field gradients. While exemplified for a Method I analysis, caution should generally be exercised. For example, the general approach put forth in Ref.~\cite{BowHobHui17}, employing Rydberg atoms, could also be susceptible to error from field gradients; this may especially be the case if the spatial sampling provided by the ballistic Rydberg atoms differs from that of the lattice-trapped clock atoms. By combining the distinct attributes of Method I (applied fields) and Method II (Faraday shielding), we have demonstrated an effective means for tackling the problem of DC Stark shifts in optical lattice clocks.

This work was supported by NIST, NASA Fundamental Physics, and DARPA QuASAR. The authors thank W.{} Zhang and D.{} R.{} Leibrandt for their careful reading of the manuscript. This work is a contribution of the National Institute of Standards and Technology/Physical Measurement Laboratory, an agency of the U.S.{} government, and is not subject to U.S.{} copyright.


%

\balancecolsandclearpage

\newcommand{\eqnmodelshift}{1}
\newcommand{\eqnshift}{2}
\newcommand{\eqnfull}{3}
\newcommand{\eqnsimple}{4}
\newcommand{\figdistro}{3}
\newcommand{\figshield}{2}
\newcommand{\figmodel}{1}
\newcommand{\tabdata}{I}
\setcounter{equation}{4}
\setcounter{figure}{3}
\setcounter{table}{1}

\begin{widetext}

\begin{center}
\textbf{Supplemental Material: Faraday-shielded, DC Stark-free optical lattice clock}
\end{center}

\section{Analytical expressions for the model}

Here we provide analytical expressions for the model described in the main text. For expressions here, we use $R=d/2$ and $L=\ell/2$ in favor of model parameters $d$ and $\ell$ of the main text. We also generalize the model somewhat by taking the charges $q_1$ and $q_2$ on the respective windows to be uniformly distributed over an area spanning a radial distance $p$ from the axis. Setting $p=R$ corresponds to charge spread out over the entire window surface, whereas $p\rightarrow0$ corresponds to a point charge on the axis. The former case is assumed in the main text. We also generalize the atomic distribution to $(2\pi)^{-3/2}s_\rho^{-2}s_z^{-1}\exp\left(-\rho^2/2s_\rho^2-z^2/2s_z^2\right)$, where $\rho$ and $z$ are the radial and axial coordinates from the center of the vacuum apparatus. The limit $s_\rho\rightarrow0$ is assumed in the main text, with the subscript on $s_z$ being omitted. The case $s_\rho=s_z$ corresponds to a spherically symmetric 3D Gaussian distribution. From the symmetry of the problem, we may write the total electric potential in the vacuum region as (Gaussian electromagnetic expressions)
\begin{equation*}
\Phi=\sum_nJ_0\left(x_{0n}\frac{\rho}{R}\right)
\left[
\alpha_n V\sinh\left(x_{0n}\frac{z}{R}\right)
+\beta_n \frac{\left(q_1-q_2\right)}{R}\sinh\left(x_{0n}\frac{z}{R}\right)
+\gamma_n\frac{\left(q_1+q_2\right)}{R}\cosh\left(x_{0n}\frac{z}{R}\right)
\right],
\end{equation*}
where $J_\nu(x)$ here and below are Bessel functions and $x_{\nu n}$ are the Bessel function zeros, $J_\nu\left(x_{\nu n}\right)=0$. The dimensionless coefficients $\alpha_n$, $\beta_n$, and $\gamma_n$ are given below in terms of the model parameters.

The clock shift is given in terms of the potential by $\delta\nu=(k/2)\left\langle\nabla^2\Phi^2\right\rangle$. Contributions to $\delta\nu$ are identified with $\delta\nu_0$, $aV$, or $bV^2$ according to their $V$-dependence. We expand $\nabla^2\Phi^2$ about the origin and use the fact that $\left\langle \rho^mz^n\right\rangle\propto s_\rho^ms_z^n$ if $m$ and $n$ are both even and $\left\langle \rho^mz^n\right\rangle=0$ otherwise. We subsequently form the combination $4(\delta\nu_0)(aV)^{-2}(bV^2)-1$, which is equivalent to $\eta$ from the main text. To leading order in $s_\rho$ and $s_z$, we find
\begin{equation*}
\eta=\frac{\zeta^2}{\mathcal{R}^2}\left(\frac{1}{4}s_\rho^2+s_z^2\right),
\end{equation*}
with $\zeta\equiv\left(q_1+q_2\right)/\left(q_1-q_2\right)$ and 
\begin{equation*}
\mathcal{R}\equiv
R\frac{\sum_n\beta_nx_{0n}}{\sum_n\gamma_nx_{0n}^2}.
\end{equation*}
Omitted terms in the expression for $\eta$ are of order $s_\rho^ms_z^n$ with $m+n\geq4$. Whereas the coefficients $\beta_n$ and $\gamma_n$ appear at lowest order, the coefficient $\alpha_n$ does not.

Solving the boundary value problem for $\Phi$, the coefficients $\alpha_n$, $\beta_n$, and $\gamma_n$ are given by
\begin{gather*}
\alpha_n=2\epsilon \frac{1}{x_{0n}}\frac{1}{J_1\left(x_{0n}\right)}
\left[
\frac{1}
{{\displaystyle\sinh\left(x_{0n}\frac{t}{R}\right)\cosh\left(x_{0n}\frac{L}{R}\right)
+\epsilon\,\sinh\left(x_{0n}\frac{L}{R}\right)\cosh\left(x_{0n}\frac{t}{R}\right)}}
\right],
\\
\beta_n=4\frac{1}{(p/R)}\frac{1}{x_{0n}^2}\frac{J_1{\displaystyle\left(x_{0n}\frac{p}{R}\right)}}{\left[J_1\left(x_{0n}\right)\right]^2}
\left[
\frac{{\displaystyle \sinh\left(x_{0n}\frac{t}{R}\right)}}
{{\displaystyle\sinh\left(x_{0n}\frac{t}{R}\right)\cosh\left(x_{0n}\frac{L}{R}\right)
+\epsilon\,\sinh\left(x_{0n}\frac{L}{R}\right)\cosh\left(x_{0n}\frac{t}{R}\right)}}
\right],
\\
\gamma_n=4\frac{1}{(p/R)}\frac{1}{x_{0n}^2}\frac{J_1{\displaystyle\left(x_{0n}\frac{p}{R}\right)}}{\left[J_1\left(x_{0n}\right)\right]^2}
\left[
\frac{{\displaystyle \sinh\left(x_{0n}\frac{t}{R}\right)}}
{{\displaystyle\sinh\left(x_{0n}\frac{t}{R}\right)\sinh\left(x_{0n}\frac{L}{R}\right)
+\epsilon\,\cosh\left(x_{0n}\frac{t}{R}\right)\cosh\left(x_{0n}\frac{L}{R}\right)}}
\right].
\end{gather*}

\newpage

\end{widetext}

\section{Clock shift functional form}

\label{Sec:validate}

For the 1D model, the simple functional form of the clock shift, Eq.~(\eqnmodelshift), can be readily inferred from the superposition principle. One envisions the following two complementary problems: (i) inclusion of the stray charge, with zero electric potential on the boundary and (ii) omission of the stray charge, with non-zero electric potential on the boundary. The field in case (i) identifies with the stray field, while the field in case (ii) identifies with the applied field. The superposition of these fields gives the total field. Since in case (ii) there are no internal charges and $V$ amounts to a common scale factor for the electric potential on the boundary ($+V$ on the top electrode, $-V$ on the bottom electrode, zero on the side walls), the applied field is consequently proportional to $V$ at all points. The first and last terms in Eq.~(\eqnmodelshift) are shifts due to the stray and applied fields, being quadratic in the respective field, while the middle term in Eq.~(\eqnmodelshift) is a cross term that is linear in both.

\begin{figure}[h]
\includegraphics[width=0.9\linewidth]{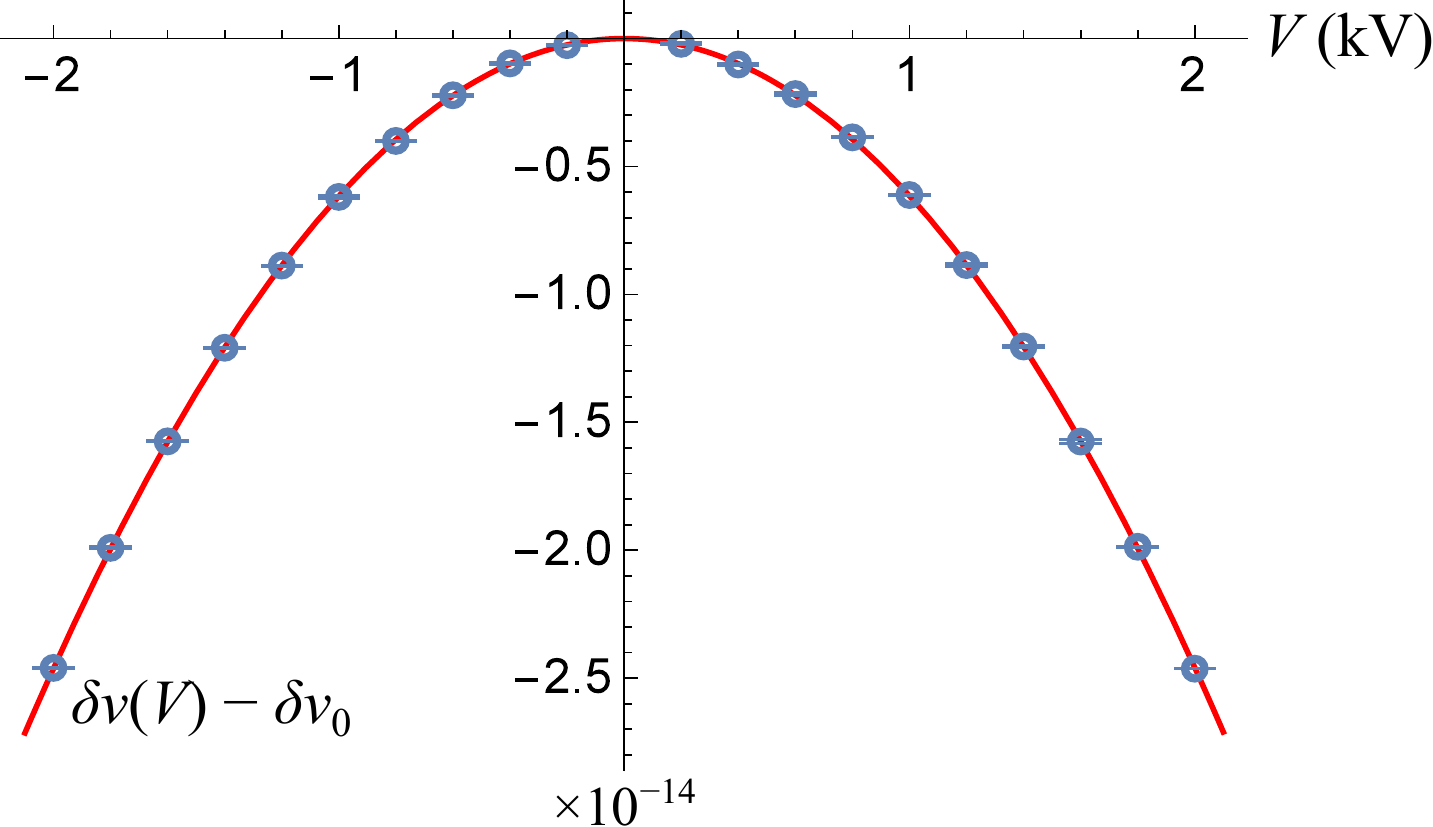}
\caption{Induced frequency shift measured relative to the grounded arrangement, $\delta\nu(V)-\delta\nu_0$. Here $V$ is the voltage applied to the top window, with the all other windows grounded. Blue circles: measured data. Red curve: fit of the data to $aV+bV^2$, with free parameters $a$ and $b$. The fit has a reduced-chi-squared $\chi^2_\mathrm{red}=1.3$.}
\label{Fig:exptcurve}
\end{figure}

While the functional form of Eq.~(\eqnmodelshift) is evident for the 1D model, we likewise expect it to hold for our Faraday shield whenever $V$ represents a uniform scale factor for the electric potential on the boundary. To demonstrate this, we take $V$ to be the voltage on the top window, with all other windows grounded. We subsequently measure the induced frequency shift with values of $V$ ranging between $-2$~kV and $+2$~kV. All measurements are relative to the $V=0$ case. The resulting data is plotted in Fig.~\ref{Fig:exptcurve}. We perform a least squares fit of this data to the functional form of Eq.~(\eqnmodelshift), $\delta\nu(V)-\delta\nu_0=aV+bV^2$, with $a$ and $b$ taken as free parameters. The resulting fit has a reduced-chi-squared $\chi^2_\mathrm{red}=1.3$, largely validating this functional form.

Equation~(\eqnshift) in the main text is a straightforward generalization of Eq.~(\eqnmodelshift), exploiting the superposition principle to accommodate multiple voltage variables.

\section{Alternate window assignment}

The shield windows are assigned voltages in terms of the variables $V_1$, $V_2$, and $V_3$ according to Fig.~\figshield{} of the main text, which is reproduced here as Fig.~\ref{Fig:shield}(A). Here we introduce an alternate assignment, presented in Fig.~\ref{Fig:shield}(B). Assignments A and B are related by a linear transformation of the variables, with the extremum $\delta\nu^*$ being invariant. It follows that $\Delta\nu$ is also invariant. For the purposes of the main text, the difference amounts to a ``bookkeeping'' choice for the data in Table~\tabdata{}. The reason for introducing assignment B will be made clear in Section~\ref{Sec:future} below.

In the following section we describe our evaluation of $\Delta\nu$ from the data in Table~\tabdata{} of the main text. The section can be read equally well from the perspective of assignment A or B. It should be understood that to apply the latter, the $V_i$ specifications given in Table~\tabdata{} must be transformed accordingly.

\begin{figure}[t]
\includegraphics[width=0.7\linewidth]{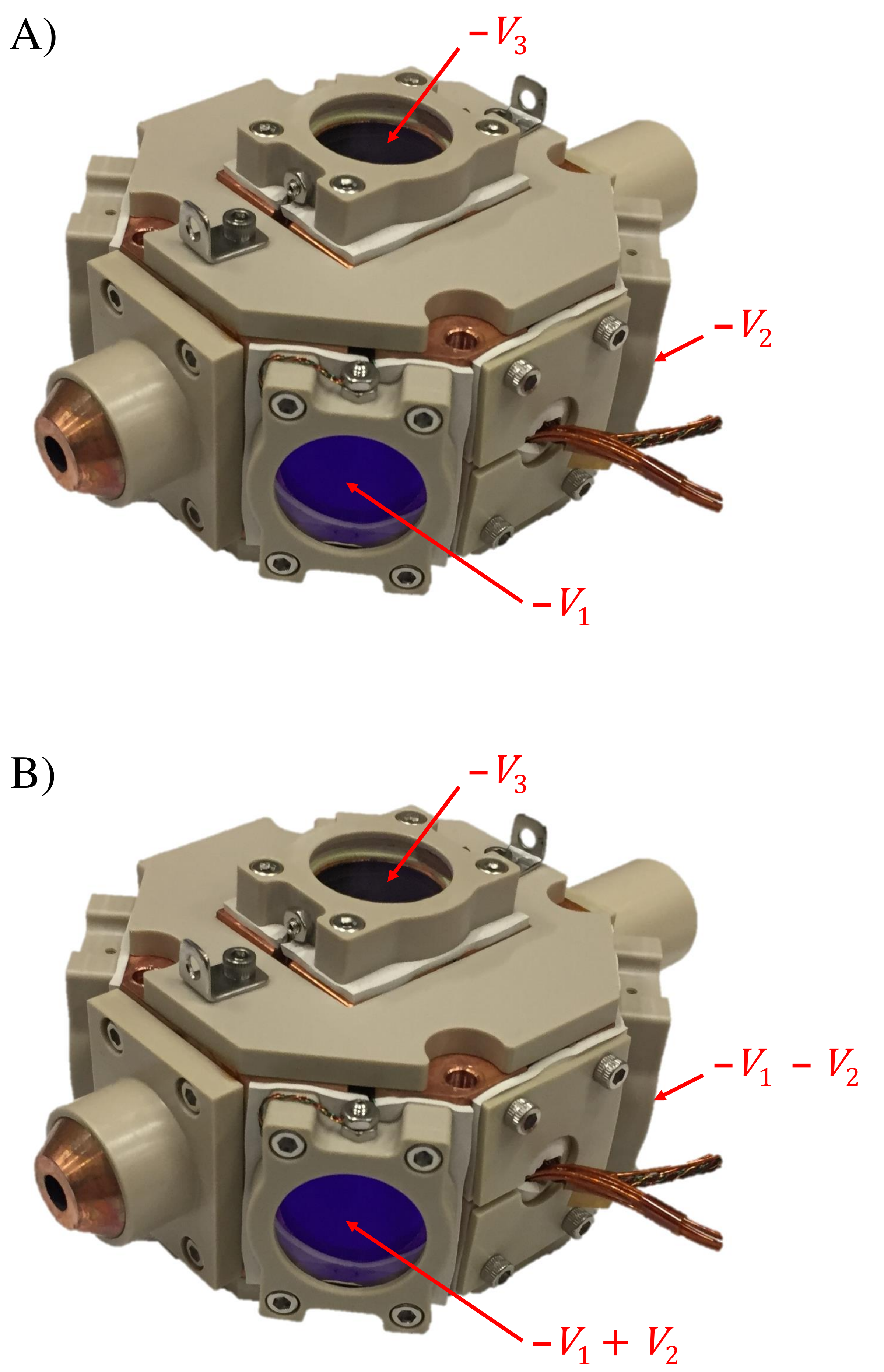}
\caption{Assignments A and B specifying the window voltages in terms of variables $V_1$, $V_2$, and $V_3$. Applied voltages are explicitly given for three windows; opposing windows have opposite voltage. For simplicity, assignment A is used in the main text. Assignment B has attributes discussed in Section~\ref{Sec:future}.}
\label{Fig:shield}
\end{figure}

\section{Monte Carlo evaluation of $\Delta\nu$}

To evaluate $\Delta\nu$ from the data in Table~\tabdata{} of the main text, we first perform a least-squares fit of Eq.~(\eqnshift) to the full data set, treating the coefficients $a_i$ and $b_{ij}$ as fit parameters (conceptually, it helps to pull $\delta\nu_0$ to the left-hand side of the equation, being that it is data for $\delta\nu\left(V_1,V_2,V_3\right)-\delta\nu_0$ that is tabulated). The resulting fit has a reduced-chi-squared $\chi^2_\mathrm{red}=1.10$, with the fit coefficients and their uncertainties given in Table~\ref{Tab:coeffs}. Equation~(\eqnfull) could then be used to relate the coefficients to $\Delta\nu$. However, propagating uncertainty to $\Delta\nu$ is complicated by correlated and non-linear variations of the coefficients. To fully respect these intricacies, we use the following Monte Carlo procedure to map probability distributions for the original data into a probability distribution for $\Delta\nu$. 1) For each data point in Table~I, a random value is pulled from the Gaussian distribution that it represents. An artificial data set is then generated by displacing the data points to these random values, while preserving error bars. 2) A least-squares fit of Eq.~(\eqnshift) to the artificial data set yields the coefficients $a_i$ and $b_{ij}$. Only the best-fit values are taken, with fit uncertainties being disregarded. 3) The coefficients are used in Eq.~(\eqnfull) to calculate $\Delta\nu$. Steps 1--3 are repeated, with the artificial data set being randomly generated each time. The observed scatter in $a_i$, $b_{ij}$, and $\Delta\nu$ are identified with probability distributions for the respective quantities. For each of the coefficients, the distribution is well-described by a Gaussian functional form with mean and standard deviation in excellent agreement with the respective entry in Table~\ref{Tab:coeffs}. The distribution for $\Delta\nu$ is presented in Fig.~\figdistro{} of the main text.


\begin{table}[t]
\caption{Coefficients $a_i$ and $b_{ij}$ derived from data in Table~I of the main text. Columns A and B are identified with respective voltage assignments in Fig.~\ref{Fig:shield}. The units are $\left(\text{clock frequency}\right)\times10^{-16}\text{/kV}^{n}$, where $n=1$ for the $a_i$ and $n=2$ for the $b_{ij}$.}
\label{Tab:coeffs}
\begin{ruledtabular}
\begin{tabular}{ccc}
& A	& B \\
\hline
\vspace{-3mm}\\
$a_1$				& $-0.011(16)$		& $-0.014(22)$			\\
$a_2$				& $-0.003(16)$		& $0.007(23)$			\\
$a_3$				& $0.10(13)$		& $0.10(13)$			\\
$b_{11}$			& $-0.715(9)$		& $-1.547(15)$			\\
$b_{22}$			& $-0.704(9)$		& $-1.291(16)$			\\
$b_{33}$			& $-228.85(7)$		& $-228.85(7)$			\\
$b_{12}$			& $-0.064(6)$		& $0.011(15)$			\\
$b_{13}$			& $-0.33(12)$		& $-0.82(17)$			\\
$b_{23}$			& $-0.49(12)$		& $-0.16(17)$			
\end{tabular}
\end{ruledtabular}
\end{table}


\section{Future analyses}

\label{Sec:future}

We may desire to reassess the DC Stark shift in the future, either periodically or after some specific event (e.g., breaking vacuum). Acknowledging the symmetry $b_{ji}=b_{ij}$, there are nine coefficients between the $a_i$ and the $b_{ij}$. This implies a minimum of nine frequency measurements to evaluate $\Delta\nu$. 
Of the nine coefficients, however, only the $a_i$ depend on the stray field. So long as there has been no significant alteration to the geometry of the shield or distribution of lattice-trapped atoms within the shield, the $b_{ij}$ therefore remain fixed. As such, we can expedite future analyses by exploiting data from the present work to specify the $b_{ij}$. Three frequency measurements will be necessary to reassess the three $a_i$.

Briefly, we envision measurement of the $a_i$. For a given $i$, we set $V_{j\neq i}=0$ and use the result
\begin{equation*}
a_i=\frac{\left(\delta\nu_+-\delta\nu_-\right)}{2\mathcal{V}},
\end{equation*}
where $\delta\nu_\pm$ denotes the clock shift under the opposite polarity conditions $V_i=\pm\mathcal{V}$. Note that there is no need to implement the fully-grounded arrangement as a reference, as the quantity in parenthesis can be obtained from a single frequency difference measurement.

As a potential simplification for future analyses, we further consider applicability of Eq.~(\eqnsimple) from the main text. Starting from the general Eq.~(\eqnfull), we expand to first order in the off-diagonal $b_{ij}$ to arrive at
\begin{equation}
\Delta\nu\approx\sum_i\Delta\nu_i-\sum_{i\neq j}\mathrm{sign}(a_ia_j)\sqrt{\Delta\nu_i}\sqrt{\Delta\nu_j}\cos\theta_{ij},
\label{Eq:firstorder}
\end{equation}
with $\Delta\nu_i\equiv a_i^2/4b_{ii}$ and $\cos\theta_{ij}\equiv b_{ij}/\left(\sqrt{b_{ii}}\sqrt{b_{jj}}\right)$.
The $\mathrm{sign}$ function appearing in Eq.~(\ref{Eq:firstorder}) returns $\pm1$ according to the sign of the argument. We note the distinction $\sqrt{x}\sqrt{y}\neq\sqrt{xy}$; namely, for $x$ and $y$ both negative, $\sqrt{x}\sqrt{y}$ evaluates to $-\sqrt{xy}$. The $\Delta\nu_i$ and the $b_{ii}$ must all be of the same sign (the sign of $k$), which excludes imaginary results for the product $\sqrt{\Delta\nu_i}\sqrt{\Delta\nu_j}$ and for the $\cos\theta_{ij}$. The $\cos\theta_{ij}$ depend on the coefficients $b_{ij}$ and satisfy the physical relation
\begin{equation*}
\cos\theta_{ij}=\frac{\left\langle\mathbf{E}_i\cdot\mathbf{E}_j\right\rangle}
{\sqrt{\left\langle E_i^2\right\rangle\left\langle E_j^2\right\rangle}},
\end{equation*}
where we assume positive $V_i$ and $V_j$ for interpretation of the right-hand-side. For uniform fields, the parameter $\theta_{ij}$ is readily identified with the angle between $\mathbf{E}_i$ and $\mathbf{E}_j$. More generally, $\cos\theta_{ij}$ incorporates an average over the atoms. In the limit that the off-diagonal $\cos\theta_{ij}$ approach zero, $\Delta\nu$ is given exactly by $\Delta\nu=\sum_i\Delta\nu_i$, which is Eq.~(\eqnsimple). Equation~(\ref{Eq:firstorder}) provides a means to assess validity of this simple relation on a case-to-case basis.

\begin{table}[t]
\caption{Off-diagonal $\cos\theta_{ij}$. Columns A and B are identified with respective voltage assignments in Fig.~\ref{Fig:shield}.}
\label{Tab:cos}
\begin{ruledtabular}
\begin{tabular}{ccc}
& A	& B \\
\hline
\vspace{-3mm}\\
$\cos\theta_{12}$	& $0.090(8)$		& $-0.008(10)$	\\
$\cos\theta_{13}$	& $0.026(9)$		& $0.044(9)$	\\
$\cos\theta_{23}$	& $0.039(9)$		& $0.010(10)$
\end{tabular}
\end{ruledtabular}
\end{table}

We use the experimental data in Table~I of the main text and the Monte Carlo technique discussed above to evaluate probability distributions for the off-diagonal $\cos\theta_{ij}$. In each case, the distribution is well-described by a Gaussian functional form. The results (mean and standard deviation) are presented in Table~\ref{Tab:cos}.

Noting the technical limitation of $\pm2$~kV on any window, we find advantage in using assignment B for future analyses. For one, $\left|\mathbf{E}_1\right|$ and $\left|\mathbf{E}_2\right|$ can be made a factor of $\sim\!\sqrt{2}$ larger for B compared to A. Ultimately this implies a factor $\sim\!2$ tighter constraint on $\Delta\nu_1$ and $\Delta\nu_2$ for a given averaging time. In either case, $\left|\mathbf{E}_3\right|$ can be made much larger than $\left|\mathbf{E}_1\right|$ and $\left|\mathbf{E}_2\right|$, as the atoms are in closer proximity to the vertical windows than the horizontal windows. Consequently, $\Delta\nu_3$ can be constrained much tighter than $\Delta\nu_1$ and $\Delta\nu_2$ for a given averaging time. With $\Delta\nu_3$ tightly constrained (and presumably tightly constrained with respect to zero), the corrective terms of principal concern in Eq.~(\ref{Eq:firstorder}) scale as $\sqrt{\Delta\nu_1}\sqrt{\Delta\nu_2}\cos\theta_{12}$. Inspecting Table~\ref{Tab:cos}, we see that $\cos\theta_{12}$ is suppressed by an order of magnitude for B compared to A. This is a result of the shield geometry. In particular, the internal copper surface is largely symmetric with respect to a vertical plane that passes through the center of the shield (where the atoms reside) and is normal to the axis defined by the two apertures. We conclude that assignment B is favorable for future analyses of the DC Stark shift, and we anticipate that the simple expression $\Delta\nu=\sum_i\Delta\nu_i$ will suffice for such analyses. By taking three frequency difference measurements $(\delta\nu_+-\delta\nu_-)$, averaged to a statistical uncertainty of $1\times10^{-17}$ for $i=1,2$ and $1\times10^{-16}$ for $i=3$, and applying the same conservative distribution for $\eta$ as in the main text, the stray-field shift can be constrained to within a few $10^{-19}$ at 95.5\% confidence. With the stability afforded by our Yb clocks, this represents less than an hour of total measurement time.

\end{document}